\documentclass[preprints,article,accept,moreauthors,pdftex]{mdpi} 

\firstpage{1} 
\pubvolume{xx}
\issuenum{1}
\articlenumber{5}
\pubyear{2019}
\copyrightyear{2019}
\history{Received: date; Accepted: date; Published: date}

\continuouspages{yes}

\pdfoutput=1

\Title{Fisher-Shannon complexity analysis of high-frequency urban wind speed time series}

\Author{Fabian Guignard $^{1,}$*, Dasaraden Mauree $^{2}$, Michele Lovallo $^{3}$, Mikhail Kanevski $^{1}$ and Luciano Telesca $^{4}$}

\AuthorNames{Fabian Guignard, Dasaraden Mauree,  Michele Lovallo,  Mikhail Kanevski and  Luciano Telesca}

\address{%
$^{1}$ \quad IDYST, Faculty of Geosciences and Environment, University of Lausanne, Switzerland.\\
$^{2}$ \quad Solar Energy and Building Physics Laboratory, Ecole Polytechnique Fédérale de Lausanne, Switzerland.\\
$^{3}$ \quad Agenzia Regionale per la Protezione dell' Ambiente di Basilicata, Potenza, Italy.\\
$^{4}$ \quad CNR, Istituto di Metodologie per l’Analisi Ambientale, Tito (PZ), Italy.}

\corres{Correspondence: Fabian.Guignard@unil.ch}

\abstract{$1Hz$ wind time series recorded at different levels (from 1.5 to 25.5 meters) in an urban area are investigated by using the Fisher-Shannon (FS) analysis.
FS analysis is a well known method to get insight of the complex behavior of nonlinear systems, by quantifying the order/disorder properties of time series. 
Our findings reveal that the FS  complexity, defined as the product between the Fisher Information Measure and the Shannon entropy power, decreases with the height of the anemometer from the ground, suggesting a height-dependent variability in the order/disorder features of the high frequency wind speed measured in urban layouts. 
Furthermore, the correlation between the FS complexity of wind speed and the daily variance of the ambient  temperature shows similar decrease with the height of the wind sensor. Such correlation is larger for the lower anemometers, indicating that ambient temperature is an important forcing of the wind speed variability in the vicinity of the ground.}


\begin{document}


\section{Introduction}

When designing urban areas, it is fundamental to consider multiple meteorological parameters. 
In particular, the buildings and the layout of the urban spaces will strongly impact the pedestrian comfort, building energy use \cite{Mauree2017a, Mauree2018}, dispersion of air pollutants, and renewable energy potential in urban planning scenarios \cite{Perera2018} due to the wind flow around the built-up spaces. 
Therefore, there is a need to understand better the influence of the obstacles (buildings, trees, street equipment) on the airflow. 
This can only be achieved with high quality and high frequency wind data that are registered through experimental campaigns and field trips.
Classical meteorological or climatic stations do not measure the wind speeds and direction with a sufficiently high vertical resolution and high frequency. 
The BUBBLE \cite{Rotach2005} campaign in the early 2000 provided useful data for development  and generalization of new parameterization schemes. 
However, one urban configuration and such a short observation period is not enough to generalize formulations and to understand the underlying physical processes.
For instance, to  determine the momentum and heat fluxes, it is necessary to measure the vertical profiles of wind speed near buildings \cite{Mauree2017b, Jarvi2018, Santiago2007}.
Additionally,  urban configurations are often characterized by small scale turbulence \cite{Christen2009}. Only high frequency wind speed data allow to identify such  turbulence.

The complexity of the wind speed in urban areas can be related to the nonlinear interactions that take place at any timescale between the average   wind speed vertical gradient, turbulent processes, shape, size and setting of buildings, etc.\ Hence, robust statistical methodologies are necessary to better characterize the time variability of wind speed in urban areas at different heights from the ground \cite{Mauree2017d, Mauree2017c}. 
The presence of heterogeneous artificial or natural surfaces close to the ground significantly increases the complexity of the turbulent structure. As an example, the high density of vertical surfaces and the ground heating could lead to the development of thermal instabilities. 

\begin{figure}
\centering
\includegraphics[width=10cm]{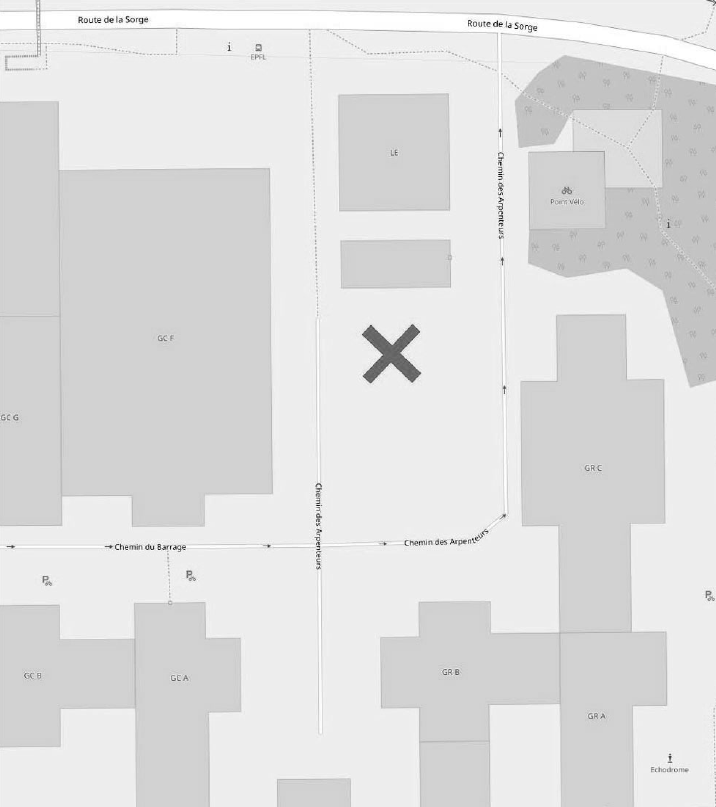}
\caption{Location of the mast.}
\label{Mast}  
\end{figure}

In this study we investigate the properties of order/disorder in the time variability of seven $1Hz$ urban wind speed time series measured at different level from the ground (from 1.5 m to 25.5 m, with 4 m spacing between each anemometer),  on a 27 m high mast located in the campus of Ecole Polytechnique Fédérale de Lausanne (EPFL), Switzerland (motus.epfl.ch). 
The average height of the building layout around the mast is about 10 m  \cite{Mauree2017d, Mauree2017c}, see Fig. \ref{Mast}.  
The experiment was performed with the aim to quantitatively evaluate how   urban buildings could impact the wind. 
Building layouts generally cause local turbulence phenomena in the wind flow below the average height of the buildings.
Thus, the main goal of the experiment was to discriminate between the turbulent dynamics of wind speed recorded by the anemometers installed  below the average building height from the “free flow” dynamics of wind speed recorded by the anemometers placed above. 
To this aim, in the present research the Fisher-Shannon method is used. 
This method jointly uses both Fisher information and Shannon entropy on time series.
Fisher-Shannon analysis has some useful applications, e.g. it allows to detect non-stationarity \cite{Vignat2003} and leads to a measure of complexity \cite{Esquivel2010}.

The paper is organized as follows. 
First, a brief description of the experiment is presented. 
Then the Fisher-Shannon method is explained.
Next, the results are discussed, and the final remarks  are summarized in the conclusions.
 
\section{Description of the experiment}

\begin{figure}
\centering
\includegraphics[width=\linewidth]{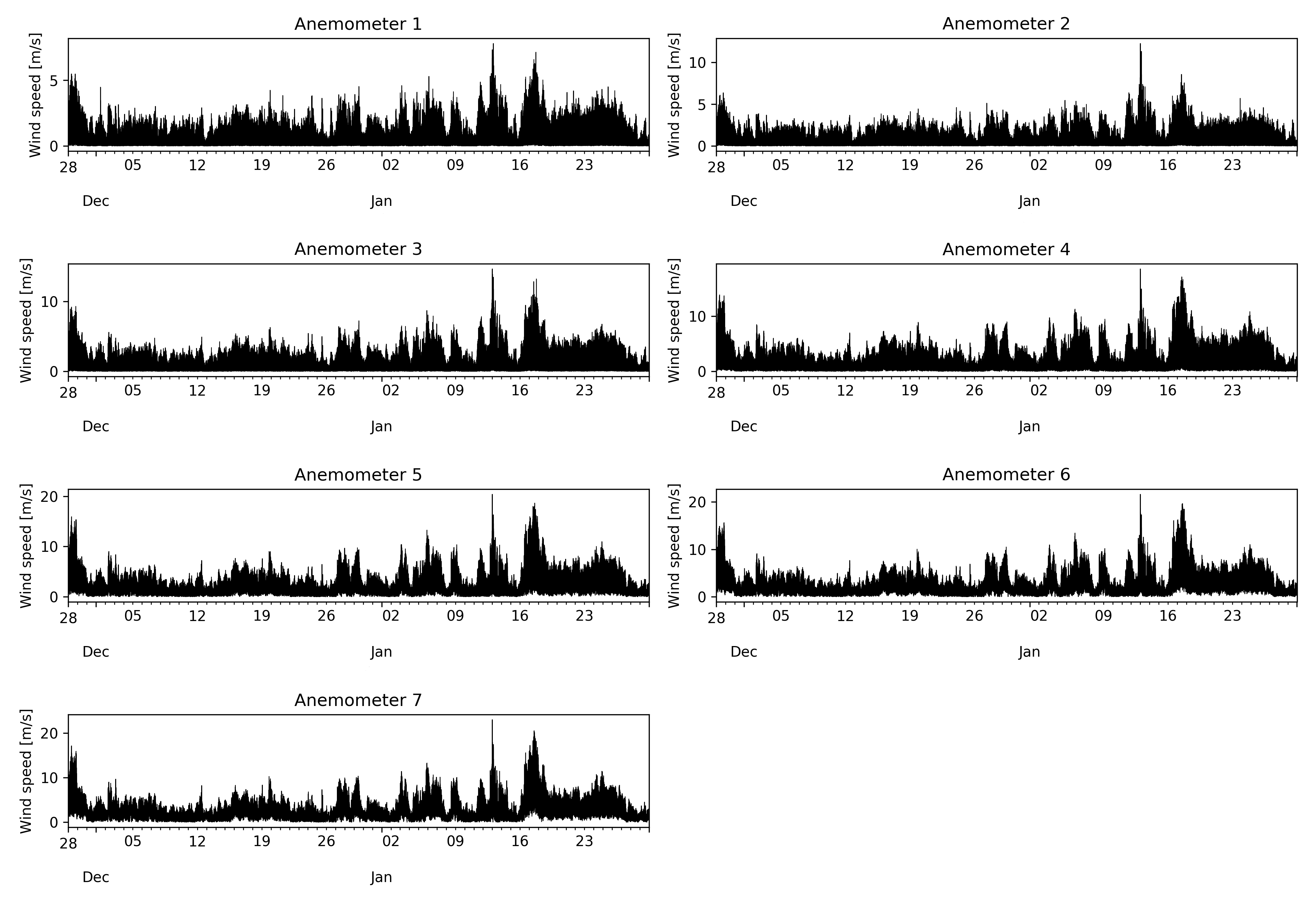}
\caption{$1 Hz$ wind speed time series for the 7 anemometers.}
\label{Data}   
\end{figure}

\begin{figure}
\centering
\includegraphics[width=\linewidth]{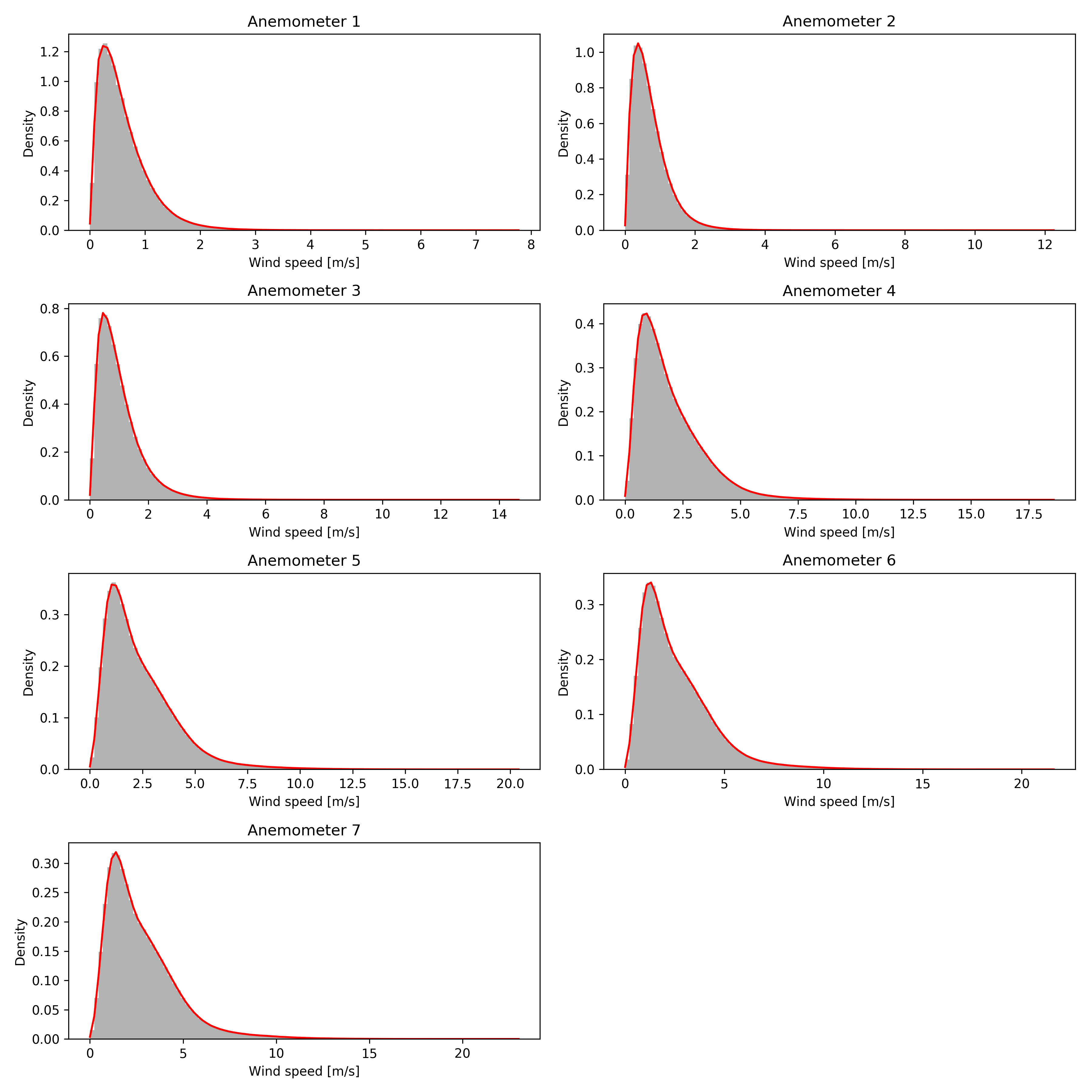}
\caption{Histograms and kernel density estimations  for the 7 anemometers.}
\label{Hist}  
\end{figure}

A meteorological mast of 27 m high, where seven sonic anemometers (Gill Wind Master) are located each 4m, has been installed in the EPFL campus.
The first anemometer is mounted at 1.5 m (corresponding to the average height of the center of an adult) while the last one is at 25.5 m above the ground. 
It can be noted that the highest sensor was placed far enough above the building layout height to be in the inertial layer  \cite{Rotach1999}.
The three velocity components, the sonic speed and temperature were measured.
The frequency of the data is $1Hz$ \cite{Kaimal1994}.
In this study the wind speed and the sonic temperature are analyzed for each level on the tower. 
The atmospheric pressure is measured at the site, with a Gill meteorological station (GMX300), also with a frequency of $1Hz$. 
The time series of the wind speed data, collected during two months period from 28 December 2016 to 29 January 2017, are shown in Fig. \ref{Data}.
Fig. \ref{Hist} shows histograms along with kernel density estimations \cite{Parzen1962, Rosenblatt1956}. Summary statistics are given in Table \ref{summary}.

\begin{table}
\centering
\begin{tabular}{lrrrrrrr}
\toprule
 &     An 1 &     An 2 &     An 3 &     An 4 &     An 5 &     An 6 &     An 7 \\
\midrule
Min.  &  0.000 &  0.000 &  0.000 &  0.000 &  0.000 &  0.000 &  0.010 \\
1st Qu. &  0.278 &  0.351 &  0.481 &  0.920  &  1.173 &  1.280 &  1.395 \\
Median & 0.493 &  0.602 &  0.824 & 1.575 &  1.965 &  2.124 &  2.295  \\
Mean &  0.606 &  0.721 &  1.009 &  1.932 &  2.388 &  2.574&  2.756  \\
3rd Qu. &  0.812 &  0.955 & 1.324 &  2.603 &  3.200 &  3.435 &  3.661 \\
Max. & 7.774  & 12.254 &  14.659 &  18.583  &  20.397 &  21.611&  23.010   \\
\bottomrule
\end{tabular}
\caption{Summary statistics of the wind speed data.}
\label{summary} 
\end{table}

\section{The Fisher-Shannon analysis}

The Fisher-Shannon (FS) method is based on the analysis of two quantities, namely, the Fisher Information Measure (FIM) and the Shannon Entropy Power (SEP).

Let $X$ be a random variable and $f(x)$ is its probability density function (pdf). FIM of $X$ is the real number $I(X)$ defined as \cite{Cover2006}
\begin{equation}
I(X) = \int_{- \infty}^{\infty} \left(\frac{\partial}{\partial x} \log f(x)\right )^2 f(x) dx. 
\end{equation}
FIM quantifies the amount of organization in the data.
SEP of $X$ is the real number, noted $N(X)$, defined as \cite{Shannon1948}
\begin{equation}
N(X) = \frac{1}{2\pi e} e^{2H(X)},
\end{equation}
where $H(X)$ is the differential entropy of $X$ given by
\begin{equation}
H(X)=-\int_{-\infty}^{+\infty}f(x)\log f(x)dx.
\end{equation}
SEP is a measure of disorder in data.

Note that FIM and SEP only depend on the pdf $f(x)$, which requires to be estimated. 
 This estimation is performed by the kernel density estimator approach \cite{Devroye1987} : for a given time series $\{x_i\}$ of length $L$,
\begin{equation}
\hat{f}_L(x)=\frac{1}{bL}\sum_{i=1}^{L}K\left(\frac{x-x_i}{b}\right),
\end{equation}
where $b$ is the bandwidth,  and $K(u)$ is the kernel function, which is assumed to be continuous, non-negative,  symmetric around zero and satisfying the following two constrains
\begin{equation}
K(0) \geq 0 \hspace{5mm} \text{and} \hspace{5mm} \int_{-\infty}^{+\infty}K(u)du=1.
\end{equation}
The computation is carried out by combining the algorithms from \cite{Troudi2008} and \cite{Raykar2006}, which use a Gaussian kernel with zero mean and unit variance :
\begin{equation}
\hat{f}_L(x)=\frac{1}{bL\sqrt{2\pi}}\sum_{i=1}^{L}e^{-\frac{1}{2}\left(\frac{x-x_i}{b}\right)^2}.
\end{equation}

Although there is no consensus about the definition of the complexity, FS analysis has been employed as a statistical complexity measure \cite{Esquivel2010, Angulo2008}. 

The FS complexity $C(X)$ is defined as the product of  FIM and  SEP,
\[
C(X) = N(X) \cdot I(X).
\]
It can be shown that $C(X) \geq 1$, with equality if and only if $X$ is a Gaussian random variable \cite{Dembo1991}, which is known as the isoperimetric inequality for entropies. Moreover, it is easy to show \cite{Vignat2003}  that for any scalar $a\in \mathbb C^*$,
\begin{align}
N(aX) &= |a|^2 N(X), \\
I(aX) &= |a|^{-2} I(X).
\end{align}
In particular, the FS complexity of a random variable is constant under scalar multiplication.

\section{Results and discussion}

\begin{figure}
\centering
\includegraphics[width=10cm]{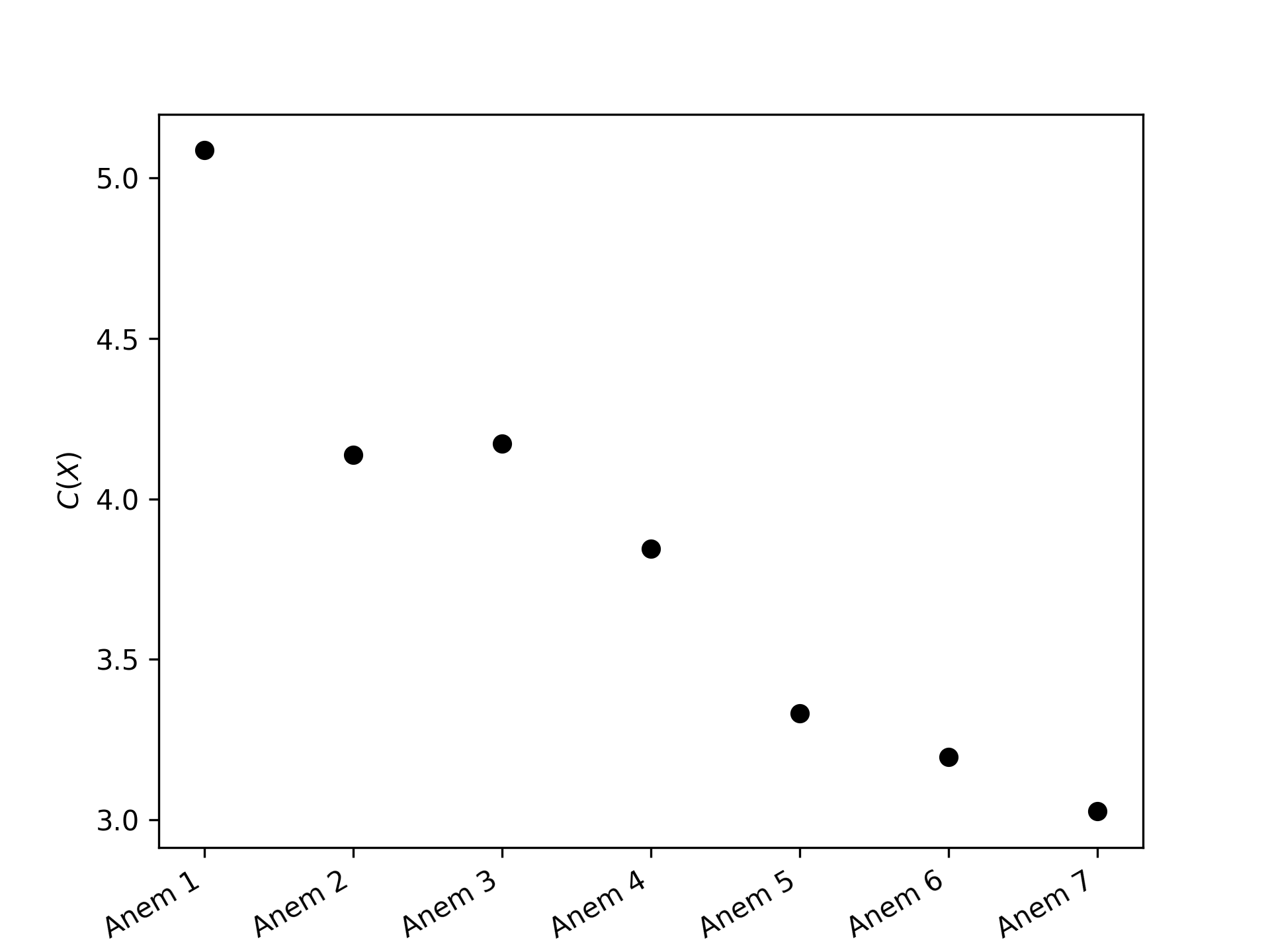}
\caption{FS complexity of the 7 wind speed time series.}
\label{Complexity} 
\end{figure}

For each series, the FIM, the SEP, and the FS complexity are calculated. 
The FS complexity for each anemometer is shown in Fig. \ref{Complexity}.
It is evident that there is a decreasing pattern of the FS complexity with height increasing from the ground.  
The largest value is reached at the anemometer placed at 1.5 m and the smallest for the one placed at 25.5 m.  

\begin{figure}
\centering
\includegraphics[width=\linewidth]{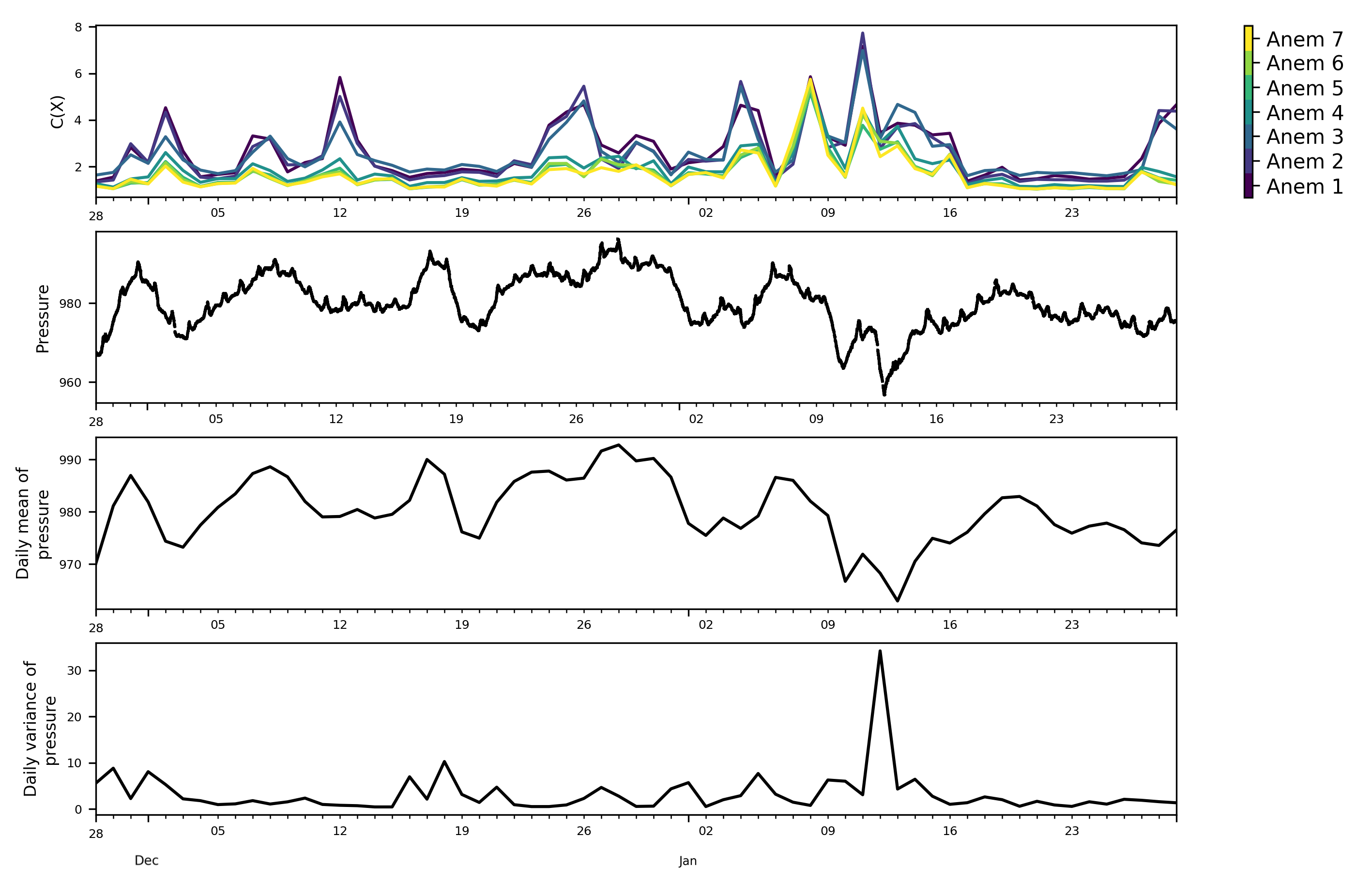}
\caption{Daily FS complexity at the 7 levels of the mast, pressure in $[hPa]$, daily mean of pressure and daily variance of pressure. }
\label{Pressure}  
\end{figure}

The daily variation of the FS complexity (top of Fig. \ref{Pressure}) reveals a clustering tendency among the seven anemometers: the lowest three anemometers are generally characterized by larger values of FS complexity during all the investigation period, while the four highest ones display smaller values of FS complexity through time. 
It is interesting to note that the lowest three anemometers are placed at height lower than the average height of the buildings surrounding the mast, while the four highest anemometers are placed well above the building average height. It is reasonable to think that such clustering of the wind series into two groups could reflect the different dynamics of the wind flow below and above the average height of the building layout \cite{Coceal2004, Christen2007, Mauree2017b, Oke2017}.
However, all curves of the daily variation of FS complexity show a certain coincidence of the occurrences of the peaks, especially in the last half of the investigation period. 

\begin{figure}
\centering
\includegraphics[width=\linewidth]{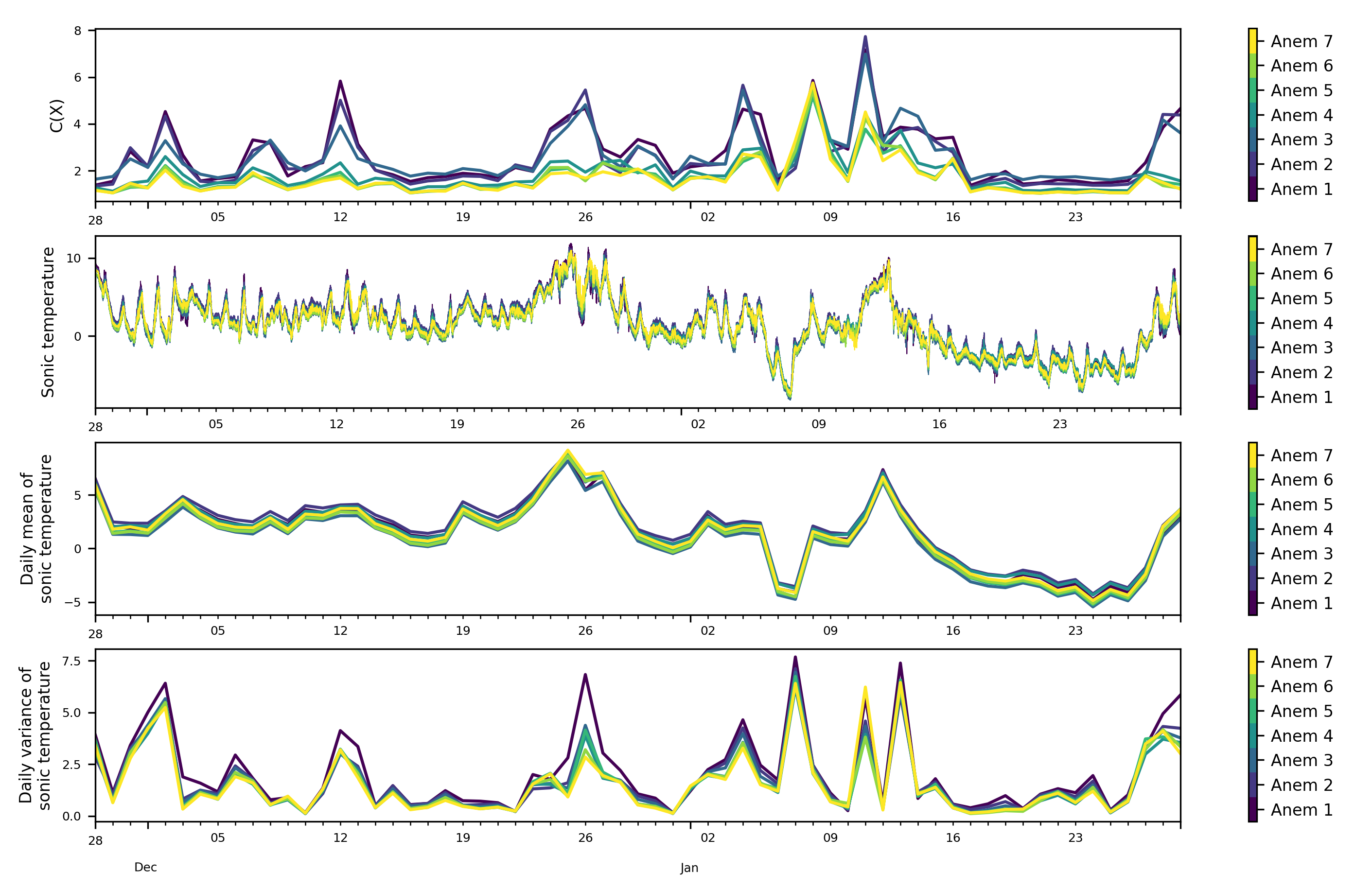}
\caption{Daily FS complexity, sonic temperature in $[ ^\circ C]$, daily mean of sonic temperature and daily variance of sonic temperature at the 7 levels of the mast.}
\label{Sonic_temp}  
\end{figure}

To see if any link could be found between the daily variations of the FS complexity with meteo-climatic variables, the daily mean and variance of the ambient pressure (Fig. \ref{Pressure}) and sonic temperature (Fig. \ref{Sonic_temp}) are calculated. 
Showing synoptically the daily variation of the FS complexity and the daily variation of the mean pressure and sonic temperature, we can observe that most of the peaks of the FS complexity seem to have a qualitative correspondence with those of the mean pressure, while no apparent correlation with the sonic temperature mean is observed.  

\begin{figure}
\centering
\includegraphics[width=10cm]{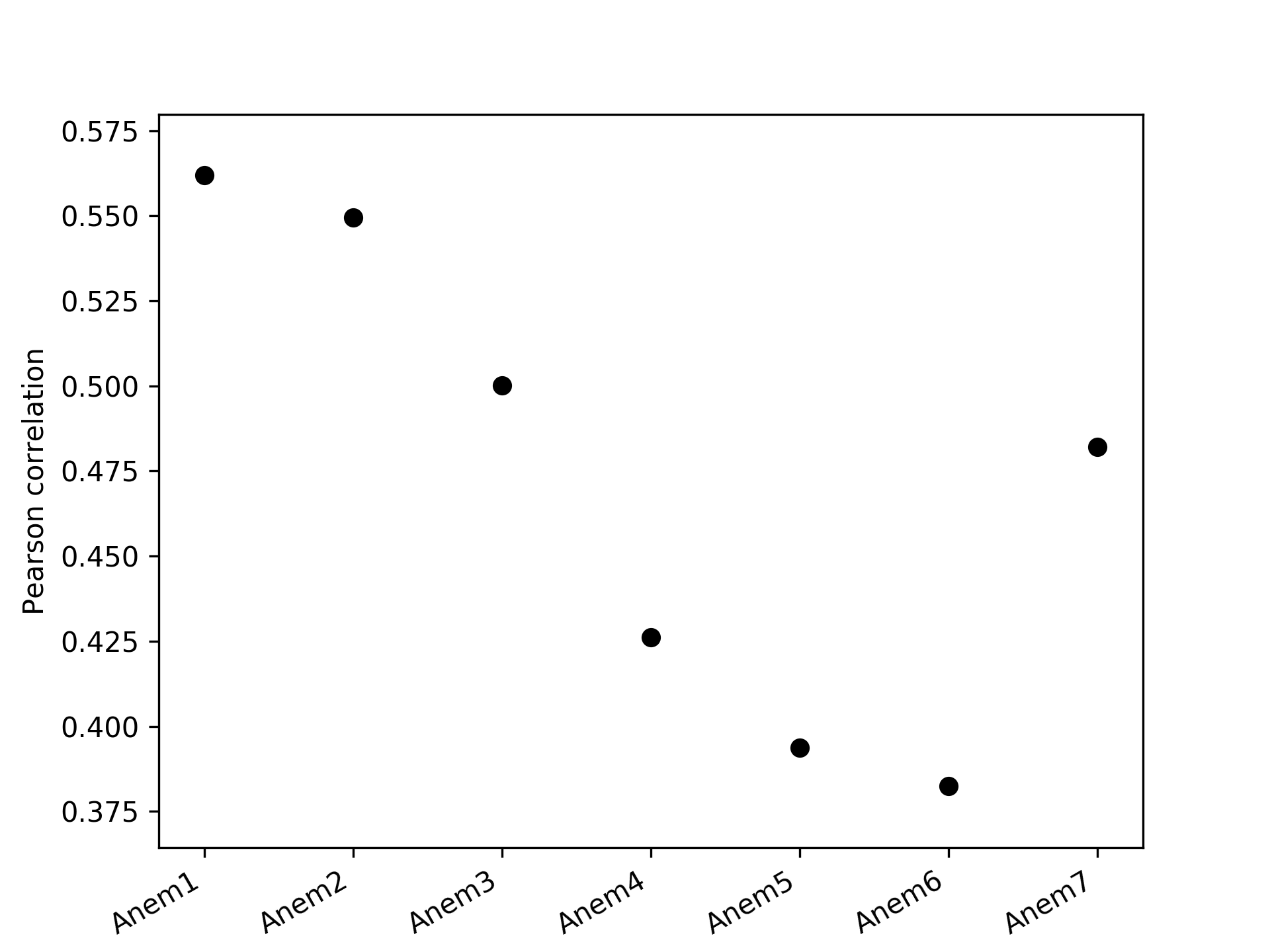}
\caption{Pearson correlation between daily FS complexity and daily variance of sonic temperature at the 7 levels of the mast. }
\label{Sonic_temp_corr}  
\end{figure}

The analysis of the correlations between the FS complexity and the variance of pressure and sonic temperature, instead, shows an apparent larger correlation between the FS complexity and the variance of sonic temperature, especially for the lower anemometers. Fig. \ref{Sonic_temp_corr} shows that the Pearson correlation coefficients between the FS complexity and the variance of the sonic temperature is larger for the lower anemometers and smaller for the higher ones. Since the data are non-normal, a non-parametric permutation test is performed for each anemometer in order to assess the significance of the correlation coefficients \cite{Davison1997}. The number of permutation is set to  $R = 999$. The results  are presented in Table \ref{permtest}.

\begin{table}
\centering
\begin{tabular}{lcc}
\toprule
 &     Correlation &    p - value \\
\midrule
Anem 1 &  0.562 &  0.001 \\
Anem 2 &  0.550 &  0.001  \\
Anem 3 &  0.500 &  0.001  \\
Anem 4 &  0.426 &  0.002 \\
Anem 5 &  0.394 &  0.002  \\
Anem 6 &  0.382 &  0.006  \\
Anem 7 &  0.482 &  0.001  \\
\bottomrule
\end{tabular}
\caption{Pearson correlation coefficient and p-value between daily FS complexity and daily variance of sonic temperature. The p-values are obtain with a non-parametric permutation test (999 permutations).}
\label{permtest} 
\end{table}

The analysis conducted in this study can suggest two driving forces. 
The atmospheric pressure obviously indicates the variation in the synoptic flows, hence gives an indication of the general weather condition during the monitoring period. 
While looking at the variance, the sonic temperature seems to demonstrate that the variation of the temperature has a non-negligible effect on the complexity found in the wind speed (at least inside the canyon). 
The impact of the radiation and the differential heating of the surfaces inside the urban canyon could also lead to the increased variances \cite{Oke2017}. 

\section{Conclusions}

An analysis of the time series of the wind speed from seven anemometers, located in an urban area on a 27 m mast at EPFL, were conducted by means of the Fisher-Shannon methods. 
The objective of this study was to determine the possible physical causes of the variability in the time series but in the vertical profile of the wind. 
In particular, the Fisher Information Measure and the Shannon Entropy Power were applied. 
The study clearly demonstrated that there was a significant decrease in the complexity with height. 
This confirmed that the presence of buildings and more generally of obstacles in the street canyon, considerably modify the wind structure and profiles in an urban setup.
Additionally, the temperatures from the anemometers as well as the atmospheric pressure from an on-site meteorological station were also used to provide data on the prevailing conditions during the monitoring campaign. 
It appears  that the atmospheric pressure could be a proxy for the synoptic flows and the current meteorological conditions. 
On the other hand, the temperature is a reflection of the amount of heat brought by general flow but also due to the heated surfaces in the canyon. 
This clearly points to the importance of new parameterization, particularly for the turbulent buoyancy term, as shown in \cite{Mauree2017b}. 
Further development will include an analysis of the atmospheric stability while taking into account also the surface temperatures of the surrounding obstacles.


\vspace{6pt} 



\authorcontributions{Software, F.G., M.L.;  formal analysis, F.G., M.L.; investigation, FG, DM, LT; data curation, F.G., D.M.; writing--original draft preparation, L.T.; writing--review and editing, F.G., D.M., M.K., L.T.; visualization, F.G.; supervision, M.K.; project administration, M.K.; funding acquisition, D.M., M.K., L.T.}

\funding{F. Guignard and M. Kanevski acknowledge the support of the National Research Programme 75 “Big Data” (PNR75) of the Swiss National Science Foundation (SNSF). 
L. Telesca thanks the support of the "Scientific Exchanges" project n$^\circ$ 180296 funded by the SNSF.
The MoTUS experiment was funded by EPFL and the ENAC Faculty and has been financially supported by the Swiss Innovation Agency Innosuisse and is part of the Swiss Competence Center for Energy Research SCCER FEEB\&D.}

\acknowledgments{The authors are grateful to Mohamed Laib, Federico Amato and Jean Golay for the profitable discussions.}

\conflictsofinterest{The authors declare no conflict of interest.} 

\abbreviations{The following abbreviations are used in this manuscript:\\

\noindent 
\begin{tabular}{@{}ll}
FS & Fisher-Shannon\\
FIM & Fisher Information Measure\\
SEP & Shannon Entropy Power\\
pdf & probability density function\\
EPFL & Ecole Polytechnique Fédérale de Lausanne
\end{tabular}}




\reftitle{References}


\externalbibliography{yes}
\bibliography{xampl}



\end{document}